# Magnetic Studies of Polycrystalline and Single-Crystal Na$_x$CoO$_2$

D. Prabhakaran[*], A.T. Boothroyd, R. Coldea and L.M. Helme

Clarendon Laboratory, Department of Physics, University of Oxford, Oxford, OX1 3PU, United Kingdom

D.A. Tennant

School of Physics and Astronomy, University of St Andrews, North Haugh, St Andrews, Fife, KY16 9SS, United Kingdom

We have investigated the preparation and bulk magnetic properties of Na$_x$CoO$_2$ in the range $0.5<x<0.95$, including those for single crystals grown by the floating-zone method in an image furnace. To characterize the samples and ascertain their quality we performed x-ray powder diffraction, electron-probe microanalysis (EPMA) and magnetization measurements. Based on our findings we are able to report conditions for the preparation of single-phase polycrystalline and single-crystal samples in the range $0.65 \leq x \leq 0.95$. For lower concentration samples, however, i.e. $x<0.65$, residual Co$_3$O$_4$ impurities above the detectable limit remained, and conditions to eliminate this have still to be identified. Investigations of the anisotropic magnetization of Na$_x$CoO$_2$ single crystals show: (i) transitions at ~320 K, ~275K and ~22 K present for all Na doping levels investigated, (ii) a strong increase in magnetization below ~8 K, (iii) a crossover in sense for the anisotropy around 8–10 K, and (iv) a hysteresis in the magnetization for certain Na doping levels at temperatures below ~15 K. Notably, neither of the high temperature anomalies (275 K, 320 K) could be observed in the powder samples.

I. INTRODUCTION

Interest in sodium cobalt oxide ($Na_xCoO_2$) has developed rapidly in recent years owing firstly to the report of a large thermoelectric power coupled with low resistivity in single crystals of $Na_{0.5}CoO_2$ (equivalently $NaCo_2O_4$),[1] and secondly to the discovery of superconductivity at temperatures below $T_c \approx 5$ K in $Na_{0.35}CoO_2$ $1.3H_2O$ formed by hydration of $Na_xCoO_2$.[2] The existence of superconductivity in a layered cobalt oxide has raised speculation about another route to high-$T_c$ superconductivity, and theoretical studies suggest an unconventional mechanism of superconductivity.[3] In addition, sodium cobalt oxide has $Na^+$ fast ion properties and is interesting as an example of a material where the magnetism and conductivity in a metallic sheet can be controlled electrochemically via intercalation.

The structure of $Na_xCoO_2$ consists of triangular $CoO_2$ sheets separated by layers of $Na^+$ ions. Metallic behaviour is achieved by doping Mott-insulating $CoO_2$ sheets with electrons donated by Na, converting magnetic $Co^{4+}$ ions ($S = ½$) into non-magnetic $Co^{3+}$ ($S = 0$). Incorporation of water between the sheets has the effect of increasing the sheet separation and decreasing the Na content. Superconductivity is observed in a narrow range of composition of the hydrated compound close to $x = 0.3$.[4]

The stability of the superconducting material is rather complex due to the mobility of the water molecules, and attempts to prepare the non-hydrated compound $Na_xCoO_2$ with $x < 0.5$ have resulted in mixed phase samples containing $Co_3O_4$ + $Na_xCoO_2$ with $x \geq 0.5$.[5] Hence, most studies have focused on the non-hydrated parent compound $Na_xCoO_2$ with $x \geq 0.5$, which shows striking thermoelectric properties, and which may provide important clues to the mechanism of superconductivity at lower doping levels. Studies of the magnetic properties are of particular interest given the possibility of an unconventional pairing mechanism. Motohashi *et al.* reported a magnetic transition at $T_m = 22$ K in the bulk susceptibility of polycrystalline $Na_{0.75}CoO_2$,[6] and the existence of static magnetic order below $T_m$ in this composition was subsequently confirmed by muon-spin rotation.[7] No magnetic order was found in polycrystalline samples of $Na_{0.65}CoO_2$.

A number of problems are encountered when preparing $Na_xCoO_2$. Several authors have reported that polycrystalline samples with $x$ close to 0.5 contain $Co_3O_4$ impurity and a higher Na content than the nominal starting composition.[8-10] The possibility of oxygen non-stoichiometry has also been recognized. Because Na is a very volatile element different techniques have been devised to control the Na content. Three such techniques are (i) reaction of the oxides with 'rapid heat-up',[9] (ii) reaction of cobalt metal with sodium hydroxide,[11] and (iii) the deintercalation of Na from compounds with a higher Na content.[11,12]

Single crystals of $Na_xCoO_2$ grown by the NaCl flux technique have been used to study the thermoelectric power,[13] and structural and transport properties.[10] These flux-grown crystals are very small, but recently Chou *et al.* succeeded in growing larger crystals of $Na_xCoO_2$ with $x = 0.71–0.74$ by the floating-zone technique.[14] The availability of sizable crystals is important for experimental techniques that require large volumes or areas of sample, such as neutron scattering. In this paper we present a detailed study of the synthesis of polycrystalline $Na_xCoO_2$ in the range $x = 0.5 – 0.95$ and the growth of single crystals in the range $x = 0.65 – 0.95$ by the floating-zone method. The effects of different preparation conditions are considered, including the use of high pressure for crystal growth. We also report measurements of the bulk magnetization of these samples, which reveal several anomalies that are present over the full range of Na content investigated. In particular we find a considerable difference in measured magnetization properties between single crystals and polycrystalline samples of the same composition. On pulverising single crystals large changes occur in the magnetization results and it is concluded that $Na_xCoO_2$ is very sensitive to particle size effects. A distinct difference is also observed between slow and rapid growth single crystals which could be due to disorder of Na.

## II. EXPERIMENTAL DETAILS

Polycrystalline $Na_xCoO_2$ ($x = 0.50–0.95$ in steps of 0.05) was prepared from high-purity (>99.98%) commercially-obtained $Na_2CO_3$ and $Co_3O_4$ by a procedure similar to the 'rapid heat-up' method.[9] The ground powders were spread on platinum foil and calcinated in air in a preheated furnace. Subsequent sintering was carried out at different temperatures (830-890°C) under flowing oxygen for 45–72 hours with intermediate grinding — see Table 1. Powder X-ray diffraction (XRD) patterns were collected after each sintering step to monitor the presence of any impurities, especially $Co_3O_4$ which could be detected to an accuracy of ~5%. Cylindrical pressed feed rods were prepared from the sintered powder for crystal growth. Crystals were grown by the floating-zone technique in a four-mirror image furnace (Crystal Systems Inc.). An atmosphere of mixed oxygen/argon pressurized up to 10 atmospheres was used during crystal growth. Crystals were grown at a speed of 3–8 mm/hr with 40 rpm counter-rotation of the feed and seed rods. Layered single crystals of size up to 40x8x2 $mm^3$ were cleaved from the as-grown crystal rod, which had a typical length of 120 mm and a diameter of 6 mm.

Polycrystalline and to a lesser extent single crystals of $Na_xCoO_2$ are sensitive to moisture in the ambient atmosphere, and so care was taken to store samples in a desiccator between experiments. The phase purity of the grown crystals was checked by powder XRD and electron-probe micro-analysis (EPMA). X-ray and neutron Laue patterns provided information on crystalline quality and alignment. Zero-field cooled (ZFC) and field-cooled (FC) magnetization measurements were performed with a commercial SQUID magnetometer in the temperature range $2<T<370$ K, and in applied magnetic fields up to 7 T.

## III. RESULTS AND DISCUSSION
### 3.1 Polycrystalline samples

Polycrystalline samples were carefully sintered in order to minimize the Na loss and consequent presence of $Co_3O_4$. The conditions used are given in the Table 1, the furnace temperature having been carefully calibrated with a standard thermocouple. The sintering conditions used by us differ slightly from those reported by other groups.[2,5,15] We found that calcination in air followed by subsequent sintering in flowing oxygen gave the best results. We also found that the product formed more rapidly if mixed powders rather than pressed pellets were used for sintering. Powders sintered at temperatures below 840°C invariably showed the presence of $Co_3O_4$ impurity for the composition range $0.5<x<0.85$ even after 4 sintering cycles. For lower Na-doped samples the success of the preparation depends sensitively on the sintering conditions, as illustrated in Table 1 for $x = 0.65$. For $x \leq$ 0.60 we were unable to prepare $Na_xCoO_2$ without the presence of small amounts of $Co_3O_4$ even after more than 6 sintering cycles under flowing oxygen. For $x > 0.60$ we were able to prepare single-phase samples (within the detectable impurity limit) using the conditions given in Table 1. However, we found that the $Co_3O_4$ impurity reappeared if single-phase samples were sintered at too high a temperature or with prolonged annealing due once again to the loss of Na.

The presence of $Co_3O_4$ shows up weakly in XRD patterns, but it can also be detected in magnetization data since a clear peak occurs at $T \approx 35$ K due to the antiferromagnetic ordering transition of $Co_3O_4$.[16] This is illustrated in Fig. 1(a), which displays the dc magnetization of a series of polycrystalline samples, some of which show the signature of $Co_3O_4$ impurity (marked with an arrow). The inset to Fig. 1(a) shows the magnetization of pure $Co_3O_4$ powder for reference. A systematic decrease in magnetization with increasing $x$ can be observed, which can be understood broadly in terms of an increase in the $Co^{4+}$ ($S = ½$) to $Co^{3+}$ ($S = 0$) ratio. This decrease, however, is not linear. To illustrate this point we plot in Fig. 1(b) the magnetization measured at 300 K as a function of Na composition. There is a clear change in behaviour near $x = 0.75$, with a much slower decrease in $M$ with $x$ above this doping level. We also observed that the magnetization of samples with a given nominal Na content varied in the temperature range below ~10 K depending on the sintering conditions (see Fig. 5 later). This may be the result of slightly different Na or oxygen contents.

For later reference we note that no anomaly can be observed in the dc magnetization of the polycrystalline samples in the vicinity of room temperature (280-320K), and we also highlight two features present in the magnetization at lower temperatures. These are (1) a sharp upturn below ~6 K, and (2) irreversible magnetic behaviour below $T_m \approx 22$ K. The latter is illustrated in Fig. 1(c), which shows the separation below $T_m$ of the field-cooled (FC) and zero-field-cooled (ZFC) magnetization of samples with $x = 0.70$ (main frame) and $x = 0.85$ (inset). There is no difference between the FC and ZFC data in the range $T_m$ to 370 K. These features are consistent with data reported by Motohashi *et al.* on a sample with $x = 0.75$.[6] The phase below $T_m$ has been studied by muon-spin rotation and a spin density wave ordering with a small ordered moment has been suggested.[7,17] The irreversibility is most striking for the highest doping level ($x = 0.95$), which exhibits a sharp drop in the ZFC signal below $T_m$ as shown in Fig. 1(a).

All the measurements just described on powder samples were performed immediately after final sintering process. The same samples were re-measured after a few weeks and differences were observed. In particular, the low temperature magnetization increased typically by 2–3 times compared to the first measurement, even though the samples had been stored in a dry environment in the intervening period. We did not detect any $Co_3O_4$ impurity in the re-measured sample, and the appearances of the particles did not changed visibly.

**3.2 Single crystal samples**

As mentioned earlier, the floating-zone crystal growth was performed in a high pressure atmosphere (up to 10 atm). This was found to be advantageous in several respects: (a) to reduce the Na evaporation, (b) to increase the convective flow for uniform mixing, and (c) to stabilize the molten-zone. The crystal rods typically contained 2 or 3 plate-like crystals several mm thick with the large faces parallel to the growth direction. Neutron Laue diffraction measurements confirmed that the crystallographic $c$ axis is perpendicular to the large faces. The largest crystals tend to have a mosaic distribution of several degrees. Despite many attempts, including the use of a crystal seed, we have so far been unable to obtain single crystals thicker than 2–3 mm. Details of the specific techniques used during the crystal growth process will be described elsewhere.[18]

Electron microscope images combined with EPMA revealed small inclusions of $Co_3O_4$ on the surfaces (a few per cent of the area) of the $x = 0.65$ and $x = 0.7$ crystals, but not on any of the higher Na-doped crystals. In addition, the lower doped crystals ($x < 0.75$) showed some variation (~10%) in the Na content from point to point on the surface, whereas the crystals with higher doping ($x > 0.75$) were found to have very little variation in the Na:Co ratio over the surface. The variation in Na content found in the analysis could be an intrinsic function of the doping level, or it could be because the lower doped crystals are more sensitive to atmospheric moisture, which reacts with the surface very quickly and may affect the measured Na:Co ratio. The average Na:Co ratio for the crystals found by EPMA did not always agree closely with the nominal value, but it was neither systematically high nor systematically low. These discrepancies, which in the worst cases were up to 20%, are difficult to understand given that the impurity content was very low. It is possible that the final Na content of the crystal is sensitive to the speed at which the crystal is grown or the growth atmosphere.

The magnetization data for single crystals differ considerably from those of polycrystalline samples of the same nominal compositions, and also show variations depending on growth conditions. To illustrate these differences we compare in Figs. 2(a) and 2(b) the temperature dependence of the dc magnetization of crystals of nominal composition $Na_{0.7}CoO_2$ grown with a fast growth rate (7–10 mm/hr and 7.5 atm pressure, Fig. 2(a)), and a slow growth rate (2–4 mm/hr and 9.5 atm pressure, Fig. 2(b)). We anticipate that not only may the slow-grown crystal (Fig. 2(b)) have a slightly lower Na content than the fast-grown crystal (Fig. 2(a)). Magnetization measurements on a polycrystalline sample of the same composition are also shown. All the data in Figs. 2 were obtained with an applied field of 100 Oe, and the crystals were oriented either with the field parallel ($H \parallel ab$) or perpendicular ($H \parallel c$) to the $ab$ plane.

There are five points to register about the data in Figs. 2(a) and 2(b). First, the magnetization (normalized to unit mass) of the powder sample is significantly lower than that of the crystals. This is consistent with data reported by Rivadulla *et al.* which also show such large differences between single crystal and polycrystalline samples this time for a doping level $x = 0.57$.[10] Second, the magnetic transition at $T_m$ = 22 K is more evident in the magnetization of the crystals than the powder, and is particularly prominent in the slow-grown crystal (Fig. 2(b)), for which the FC and ZFC curves are seen to split sharply below $T_m$. Third, the magnetization is anisotropic, with $M_{ab} > M_c$ over most of the temperature range as observed previously.[14,19] However, below 5–8 K the anisotropy switches sense so that $M_c > M_{ab}$ — see inset to Fig. 2(b). Moreover, the transition at $T_m$ is much clearer with $H \parallel c$ than with $H \parallel ab$. Fourth, the magnetization curves for the crystals exhibit two high temperature anomalies (both field orientations), one at $T_1 \approx 275$ K and the other at $T_2 \approx 320$ K.

Below $T_2$ there is a sudden splitting between the FC and ZFC curves, and at $T_1$ there is a change in the slope. Neither of these anomalies are visible in the powder data. Fifth, and finally, there is a steep rise in magnetization below ~ 8 K.

Broadly speaking, the five features just described are also evident in the magnetization data we have collected for crystals with other Na doping levels. Figures 3(a)–(c) shows similar curves to Figs 2 but for $x = 0.65$ (inset to Fig. 3(a)), $x = 0.75$, $x = 0.80$ and $x = 0.85$. Overall, the magnetization is seen to decrease with increasing doping, consistent with the powder measurements shown in Fig. 1(a). However, the conditions under which the crystals are grown can have a strong influence on this. For example, in Fig. 3(b) we show magnetization data measured with $H \parallel c$ for two crystals of $x = 0.80$, both grown under 10 atm applied pressure but one at a scanning rate of 4 mm/hr (crystal A) and the other at 6 mm/hr (crystal B). Over most of the temperature range the magnetization for crystal A is almost a factor two larger than crystal B, and the separation between FC and ZFC curves is also larger for crystal A than crystal B. Interestingly, the FC curve for crystal A exhibits steps at 28 K and at 10 K in addition to that at $T_m = 22$ K (inset to Fig. 3(b)).

In the case of $x = 0.85$ (Fig. 3(c)) we have included an additional magnetization curve. This was obtained from one of the seed rods prior to crystal growth. The seed rod was formed by compression of the single-phase sintered powder in a hydrostatic press, followed by an anneal at 900$^{\circ}$C in flowing oxygen. The magnetization of the seed rod material shows a broad peak centred near 50 K, similar to that exhibited by $Co_3O_4$ (inset to Fig. 1(a)), but shifted up in temperature by ~10 K. During the anneal we observed a small weight loss. These observation suggest that evaporation of Na occurred during annealing, allowing the formation of ~10 % by mass of $Co_3O_4$ impurity. However, it is interesting that this impurity disappeared during crystal growth from the same seed rod, as can be seen from the crystal data in Fig. 3(c). The starting powder used for the crystal growth was prepared under slightly different conditions to the material used in the polycrystalline studies described in section 3.1. This is why the powder magnetization data in Fig. 3(c) differs slightly from that in Fig.1(b).

The magnetization data for the $x = 0.80$ and 0.85 crystals contain anomalies at low temperatures in addition to the magnetic transition at $T_m = 22$ K. The FC curve for crystal A ($x = 0.80$) exhibits steps at 28 K and at 10 K (inset to Fig. 3(b)), and all the curves for the $x = 0.85$ crystal have a smooth rise at low temperature.

All the magnetization measurements described so far were performed in an applied magnetic field of 100 Oe. Measurements were also made in different applied fields, either temperature scans at fixed field, or field scans at fixed temperature. Figure 4 illustrates the variation of the ratio $M(T)/H$ at low temperatures for four selected field values between 100 Oe and 10,000 Oe (0.001 to 1 Tesla). ZFC data are shown for a powder sample (Fig. 4(a)) and a single crystal (Figs. 4(b) and (c)) of $Na_{0.7}CoO_2$. The $M/H$ curves for the powder sample are virtually independent of applied field, and for $T > T_m$ the curves for the crystals are also virtually the same apart from the 100 Oe field value. The difference between the 100 Oe curves and the higher field data is very probably due to remnant fields from the shields surrounding the SQUID sensor which make the actual field at the sample differ by a few Oe from the nominal field. Below $T_m$, however, the shape of $M/H$ vs $T$ measured on the crystal changes dramatically with field. The rapid increase in signal as $T \to 0$ is suppressed with field, and is virtually eliminated at 10,000 Oe. The drop in the ZFC magnetization below $T_m$ becomes smaller with increasing field, but $T_m$ itself does not shift at all. For the particular value $H = 500$ Oe the ZFC magnetization actually rises below $T_m$.

Measurements of the field dependence of the magnetization at fixed temperature were made on several of the powder and single crystal samples. Figs. 5(a) and (b) show, respectively, hysteresis loops for powder and single crystal samples of $Na_{0.75}CoO_2$ measured between –500 Oe and 500 Oe. Fig. 5(a) compares results obtained on two powder samples at a temperature of 2 K. The first measurement was made on sintered powder, prepared according to the conditions given in Table 1, and the second measurement was made on material from the same batch of sintered powder after annealing it in oxygen at 890°C for a further 15 hrs. It can be seen that the annealed powder sample exhibits a clear hysteresis, whereas the sintered powder does not. Also, the annealed powder has a much larger magnetization at a given field than the sintered powder. As mentioned earlier, the low temperature magnetization of powder samples was observed to increase over a period of several weeks after preparation, and these aged powder samples also showed a large hysteresis that had not been present in the same samples when freshly prepared. Such an ageing effect was not observed for the single crystal samples. The observation of hysteresis has been reported previously,[6] but our experiments show that the presence of hysteretic behaviour is dependent on preparation conditions and ageing time.

Fig. 5(b) shows hysteresis loops for a single crystal of $Na_{0.75}CoO_2$ measured at 2 K and 20 K with $H \parallel c$. The inset shows high field data at 2 K. Hysteresis is observed in the 2 K $M(H)$ data but hardly at all in the 20 K data, confirming the conclusion of that the hysteresis is a property of the magnetic phase below $T_m$.[6] We also observed hysteretic behaviour below $T_m$ for $H \parallel ab$, and for other Na doping levels of $x = 0.65$–$0.85$ in the single crystal samples. The amount of hysteresis was found to be greatest at $x = 0.75$. These results show that the irreversible magnetic phase is not confined to the composition $Na_{0.75}CoO_2$.

Figs. 6(a) and (b) show the variation of $M/H$ with applied field extended up to 7 T (70,000 Oe) for polycrystalline and single-crystal $Na_{0.85}CoO_2$, respectively. For the crystal the field was applied parallel to the $ab$ plane, but similar results were obtained with $H \parallel c$. At 2K, the polycrystalline sample exhibits an almost linear decrease of $M/H$ with $H$, whereas the single crystal has a rapid initial fall followed by a slower decrease at high fields. At higher temperatures, the data from the polycrystalline sample is virtually constant with field, and the single-crystal data is constant apart from a slight initial drop with field between 0 and ~1 T. The higher magnetization value for the powder sample compared to the crystal, contrary to what has been described earlier, is in this case due to the ageing effect (the powder sample was left several weeks before collecting the data shown in Fig. 6(a)).

**IV Discussion and Conclusions**

In this work we investigated the preparation of $Na_xCoO_2$ in polycrystalline and single crystal form, and we have reported how the magnetic properties (as reflected in the dc magnetization) vary with Na doping and with preparation conditions. We have found that care is needed with the sintering and annealing conditions in order to limit the presence of $Co_3O_4$ impurity in polycrystalline samples, especially for $x \leq 0.60$. We have shown that it is possible to prepare relatively large crystals of $Na_xCoO_2$ by employing the floating-zone method with a high pressure atmosphere (~10 atm) to limit the evaporation of Na during the melt growth.

Our magnetization studies have revealed differences in behaviour between polycrystalline and single crystal samples. First, the magnetization of the polycrystalline samples is systematically lower than the powder-averaged magnetization of crystals of the same nominal composition, and second the single crystals all show magnetization anomalies around room temperature, which are completely absent in the polycrystalline samples. One factor could be Na evaporation, which may occur during crystal growth and lead to a reduction in the Na content of the crystals, however compositional analysis using EPMA shows this effect is far too small to account for the observed differences. Another possibility is that there may be differences in the oxygen content of the crystals compared with the corresponding powders, or in the occupancy of different crystallographic sites by Na and O between the Co–O layers. Some evidence that the oxygen content can vary comes from the large change in magnetization at low temperatures induced by annealing polycrystalline samples in oxygen — see Fig. 5(a). Also, we observed a change in magnetization with ageing over a period of several weeks in polycrystalline samples, but not in single crystals. The relatively large surface-area-to-volume ratio of the powders is assumed to be responsible for this difference.

The two anomalies we have observed around room temperature in single crystals, suggestive of two successive ordering transitions, have not been reported previously in magnetization data and warrant further investigation. The temperatures $T_1 \approx 275$ K and $T_2 \approx 320$ K at which these features are observed are virtually independent of Na content over the range studied ($x$ = 0.65 to 0.95). Anomalies near room temperature have, however, been reported in other physical properties. Gavilano *et al.* observed anomalies near 250 K and 295 K in the $^{23}$Na-NMR response from polycrystalline $Na_{0.70}CoO_2$, and concluded that $Co^{3+}/Co^{4+}$ charge ordering occurs in this temperature range.[20] However, a localization of charge within the Co–O layers would be expected to show up in the electrical transport and no corresponding features have been found in the resistivity. Another possible explanation is the ordering of Na ions among the two Na sites between the Co–O layers. Such cation ordering could potentially cause irreversibility in the magnetization (as observed below $T_2$) as it would distort the local oxygen environment of Co ions as well as conditioning the electrostatic potential, both of which can have strong influence on the magnetism of a strongly correlated electronic state, and lead to cluster glass type effects. The fact that irreversibility is observed in crystals and not powders could then be due to the relaxation of lattice constraints due to the large surface area.

The data we have collected on single crystals confirm the presence of an ordering transition below $T_m = 22$ K believed to correspond to the transition from a paramagnetic phase to a commensurate or incommensurate spin density wave (SDW) state. Previously, this transition has been reported for samples with Na doping levels of $x = 0.75$ and $x = 0.9$.[6,7,17] Our experiments have shown that this phase is present between doping levels as low as $x = 0.65$ and as high as $x = 0.95$, with $T_m$ almost constant with $x$. We also find that there is a change in the magnetic anisotropy from easy-plane ($M_{ab} > M_c$) to easy-axis ($M_c > M_{ab}$) within the SDW phase, and that the SDW transition is more pronounced in the $H \parallel c$ magnetization compared to the $H \parallel ab$ magnetization. These observations suggest that the SDW phase may have a weak ferromagnetic or canted moment along the $c$ axis, consistent with the conclusion from a muSR study that there is a $c$-axis static internal field for $T < T_m$.[17]

In summary, the magnetization of $Na_xCoO_2$ has revealed a number of intriguing features present over a range of Na doping, some of which appear very different in single crystals compared to polycrystalline samples. Investigations with other techniques, particularly diffraction techniques, are needed to explain the transitions at $T_1$ and $T_2$, and to find out the type of static order present below $T_m$. A careful comparison of the crystal structure and chemical composition of single crystal and polycrystalline $Na_xCoO_2$ would be very valuable.

**Acknowledgments**

We would like to thank Fred Wondre for technical support, and Norman Charnley of the Department of Earth Sciences, Oxford University, for performing the EPMA analysis. We are grateful to the Engineering and Physical Sciences Research Council of Great Britain for financial support.

# References

[1] I. Terasaki, Y. Sasago, and K. Uchinokura, Phys. Rev. B **56**, R12685 (1997).

[2] K. Takada, H. Sakurai, E. Takayama-Muromachi, F. Izumi, R.A. Dilanian, and T. Sasaki, Nature **422**, 53 (2003).

[3] G. Baskaran, Phys. Rev. Lett. **91**, 097003 (2003).

[4] R.E. Schaak, T. Klimczuk, M.L. Foo, and R.J. Cava, Nature **424,** 527 (2003).

[5] C. Fouassier, G. Matejka, J-M. Reau, and P. Hagenmuller, J. Solid State Chem. **6**, 532 (1973).

[6] T. Motohashi, R. Ueda, E. Naujalis, T. Tojo, I. Terasaki, T. Atake, M. Karppinen, and H. Yamauchi, Phys. Rev. B **67**, 064406 (2003).

[7] J. Sugiyama, H. Itahara, J.H. Brewer, E.J. Ansaldo, T.Motohashi, M. Karppinen, and H. Yamauchi, Phys. Rev. B **67**, 214420 (2003).

[8] Y.Ono, R.Ishikawa, Y.Miyazaki, Y.Ishii, Y. Morii, and T. Kajitani, J. Solid State Chem. **166**, 177 (2002).

[9] T.Motohashi, E. Naujalis, R. Ueda, K.Isawa, M. Karppinen, and H. Yamauchi, Appl. Phys. Lett. **79**, 1480 (2001).

[10] F. Rivadulla, J.S. Zhou, and J.B. Goodenough, cond-mat/0304455.

[11] B.L. Cushing and J.B. Wiley, J. Solid State Chem. **141**, 385 (1998).

[12] S. kikkawa, S. Miyazaki, and M. Koizumi, J. Solid State Chem. **62**, 35 (1986).

[13] I. Terasaki, Y. Sasago, and K. Uchinokura, Phys. Rev. B **56**, R12685 (1997).

[14] F.C. Chou, J.H. Cho, P.A. Lee, E.T. Abel, K. Matan, and Y.S. Lee, cond-mat/0306659.

[15] R.J. Balsys and R.L. Davis, Solid State Ionics **93**, 279 (1997).

[16] W.L. Roth, J. Phys. Chem. Solids **25**, 1 (1964).

[17] J. Sugiyama, J.H. Brewer, E.J. Ansaldo, H. Itahara, T. Tani, M. Mikami, Y. Mori, T. Sasaki, S. Hebert, and A. Maignan, cond-mat/0310516.

[18] D. Prabhakaran et al. (unpublished).

[19] Y. Wang, N.S. Rogado, R.J. Cava, and N.P. Ong, Nature **423**, 425 (2003).

[20] J.L. Gavilano, D. Rau, B. Pedrini, J. Hinderer, H.R. Ott, S. Kazakov, and J. Karpinski, cond-mat/0308383.

Table 1. Sintering conditions used in the preparation of polycrystalline Na$_x$CoO$_2$ samples.

| Starting Composition | Sintering Temperature/atmosphere/period | | | Co$_3$O$_4$ impurity |
| --- | --- | --- | --- | --- |
| | 1$^{st}$ | 2$^{nd}$ | 3$^{rd}$ | |
| x=0.5-0.6 | 840°C/air/10h | 870°C/O$_2$/15h | 890°C/O$_2$/15h | Yes |
| x=0.65 | 820°C/air/10h | 840°C/O$_2$/15h | 850°C/O$_2$/15h | Yes |
| x=0.65 | 890°C/air/10h | 900°C/O$_2$/15h | 900°C/O$_2$/15h | Yes |
| x=0.65 | 850°C/air/10h | 870°C/O$_2$/15h | 890°C/O$_2$/15h | No |
| x=0.7-0.85 | 850°C/air/10h | 870°C/O$_2$/15h | 890°C/O$_2$/15h | No |
| x=0.9-0.95 | 850°C/air/10h | 870°C/O$_2$/15h | 870°C/O$_2$/15h | No |

**Figure Captions**

Fig. 1(a). Temperature dependence of zero-field cooled (ZFC) dc magnetization of polycrystalline Na$_x$CoO$_2$ samples. The arrow indicates the presence of Co$_3$O$_4$ impurity phase in the final compound (x≤0.6). The magnetization of Co$_3$O$_4$ is shown for comparison in the insert figure.

Fig. 1(b). Dependence of the room temperature magnetization of Na$_x$CoO$_2$ powder samples on *x*.

Fig. 1(c). Temperature dependence of dc magnetization for Na$_{0.7}$CoO$_2$ powder sample showing irreversible behaviour (separation of ZFC and FC curves) below $T_m$ = 22 K. Similar behaviour is present in the *x*=0.85 sample shown in the insert figure.

Fig. 2(a). Temperature dependence of the dc magnetization of a fast-grown Na$_{0.7}$CoO$_2$ single crystal (*H*||*ab* and *H*||*c*) and powder samples. The insert figure shows the 320K transition which is observed only in the crystal and not in the powder sample.

Fig. 2(b). Temperature dependence of the dc magnetization of a slow-grown Na$_{0.7}$CoO$_2$ single crystal measured with H||ab and H||c, showing transitions at $T_m$ and around room temperature (270K and 320K). The insert figure illustrates the sharp transition at 22 K, and the steep rise below ~8 K.

Fig. 3(a). Temperature dependence of the dc magnetization of a Na$_{0.75}$CoO$_2$ single crystal measured with *H*||*ab* and *H*||*c* (ZFC and FC) in a 100 Oe applied field. The insert figure shows the presence of transitions at $T_m$, 270K and 320K in a single crystal of Na$_{0.65}$CoO$_2$.

Fig. 3(b). Temperature dependence of the dc magnetization of Na$_{0.8}$CoO$_2$ single crystals: crystal A (slow grown) and crystal B (fast grown) respectively. Insert figures shows multiple transitions below $T_m$ observed in crystal A.

Fig. 3(c). Dc magnetization versus temperature curve for Na$_{0.85}$CoO$_2$ single crystal measured with *H*||*ab* and *H*||*c* compared with different powder samples. An arrow in the insert figure shows the presence of Co$_3$O$_4$ impurity phase in the high temperature sintered powder which is not observed in the crystal.

Fig. 4. Variation of *M/H* with temperature for Na$_{0.7}$CoO$_2$ samples, (a) powder, (b) crystal with *H*||*ab*, and (c) crystal with *H*||*c*, measured at different applied fields.

Fig. 5. Dc magnetization loops for Na$_{0.75}$CoO$_2$ samples, (a) powder, and (b) crystal with *H*||*c*, measured at different temperatures. The insert figure shows the high-field magnetization loop for the crystal (*H*||*c*) measured at 2 K.

Fig. 6. Variation of *M/H* with applied field for Na$_{0.85}$CoO$_2$ samples, (a) powder and (b) crystal with *H*||*ab*, measured at different temperatures under ZFC conditions.

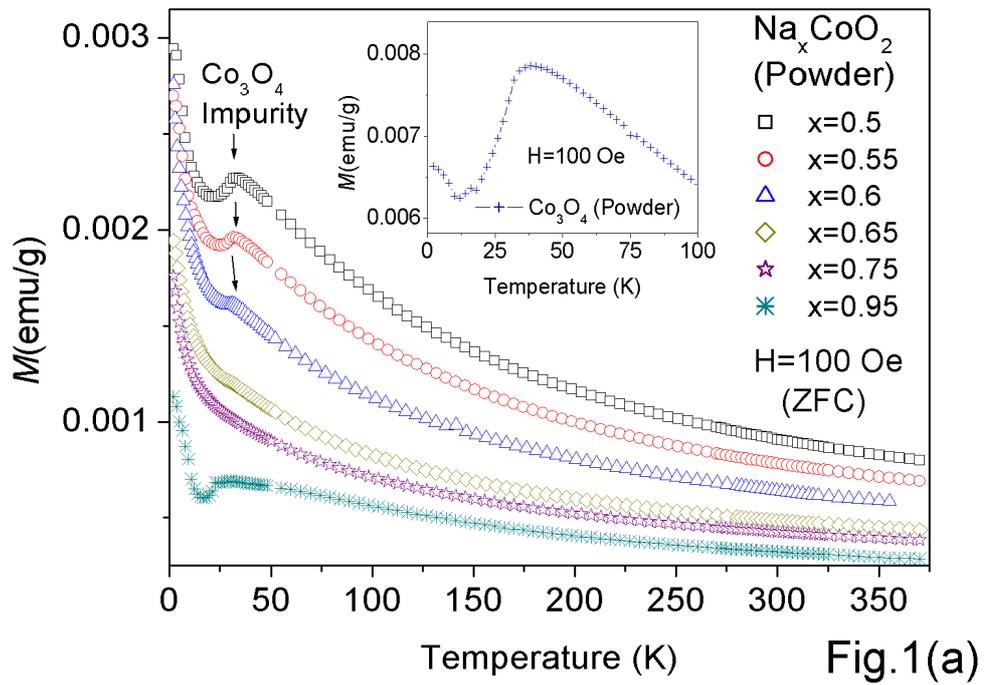

Fig.1(a)

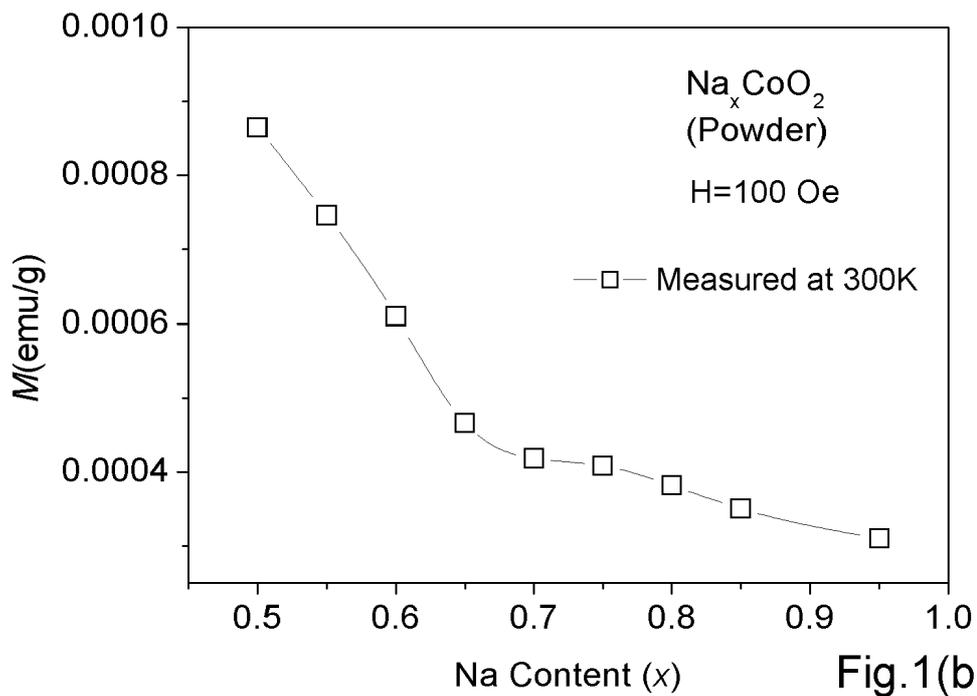

Fig.1(b)

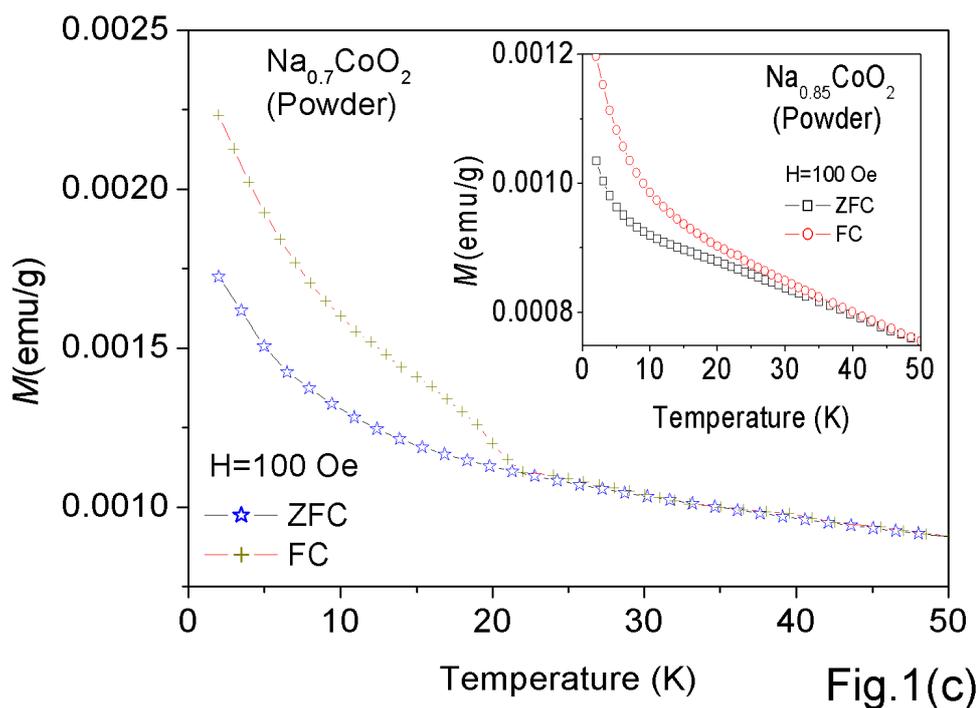

Fig.1(c)

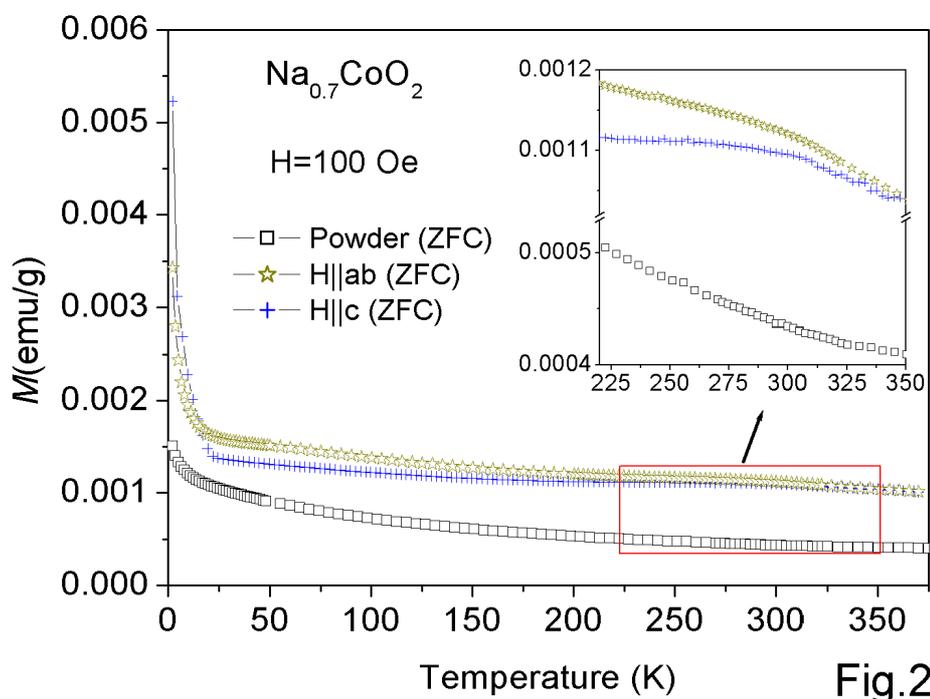

Fig.2(a)

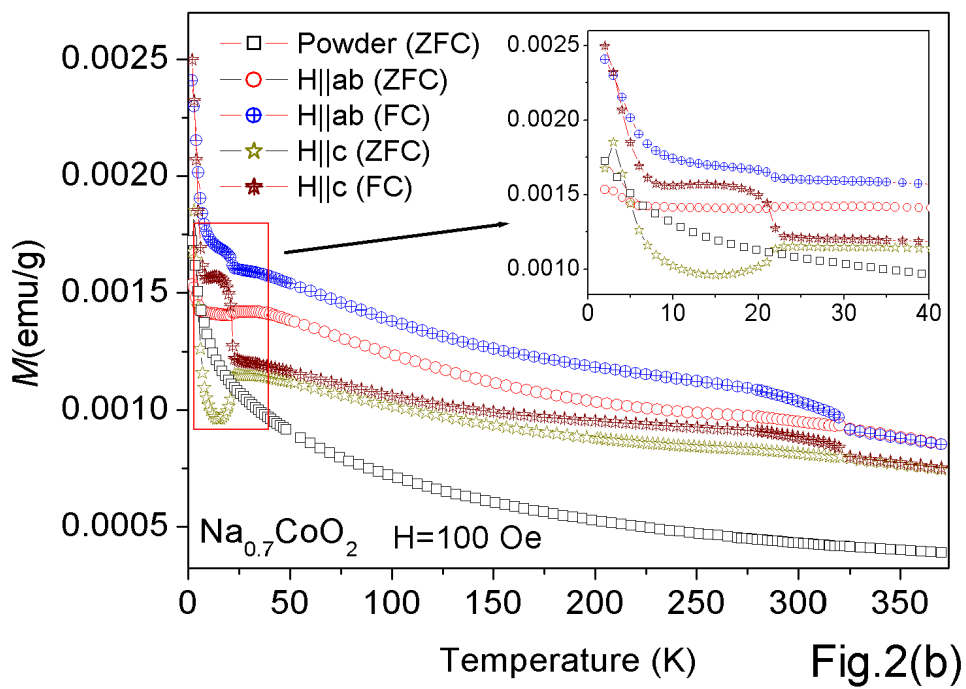

Fig.2(b)

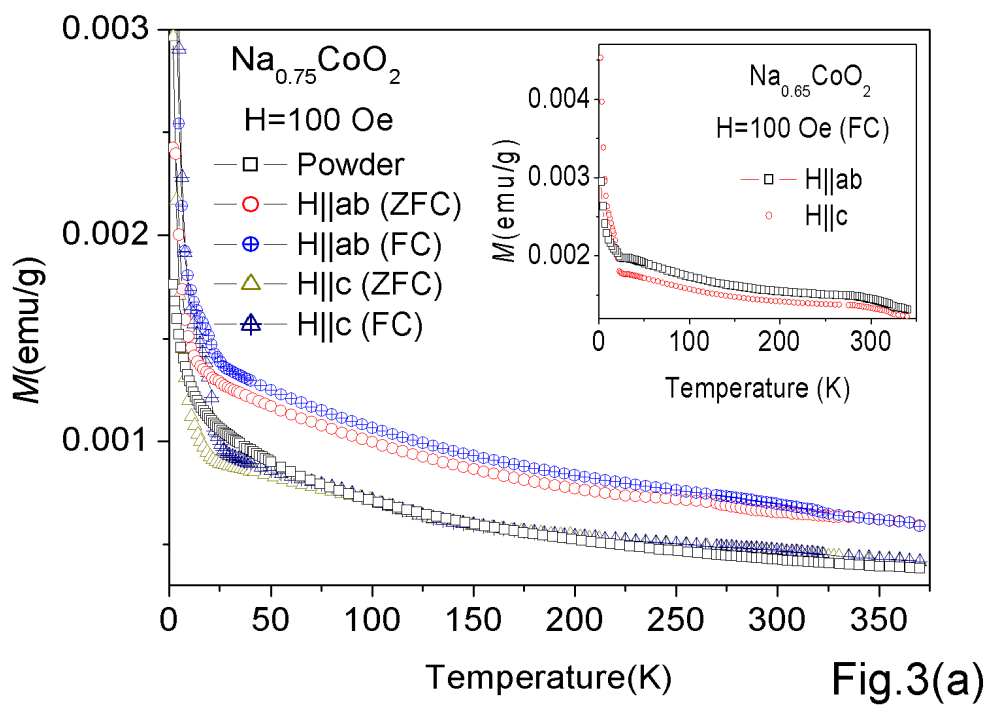

Fig.3(a)

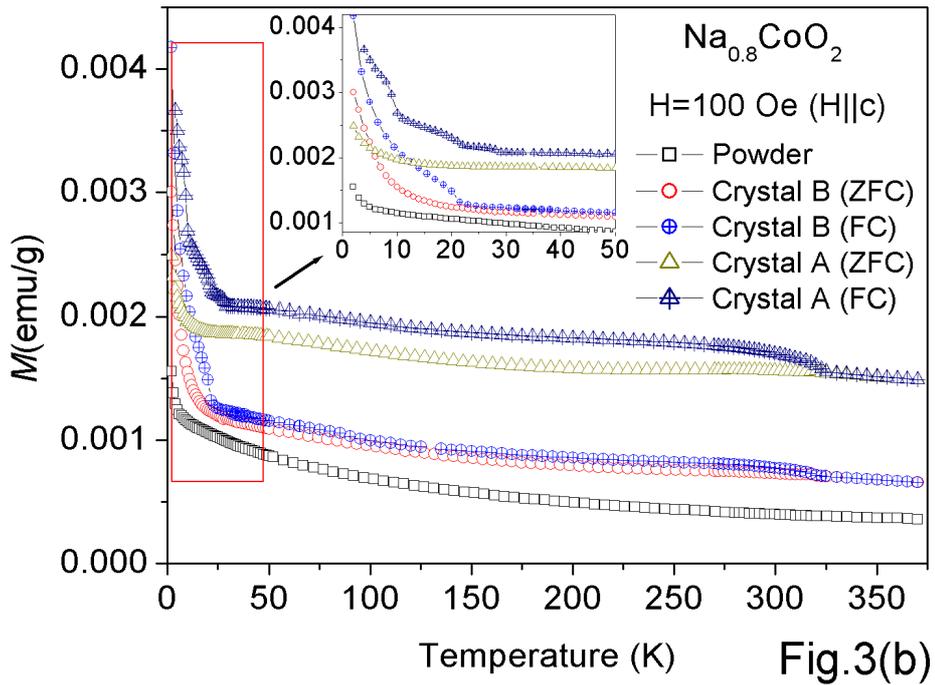

Fig.3(b)

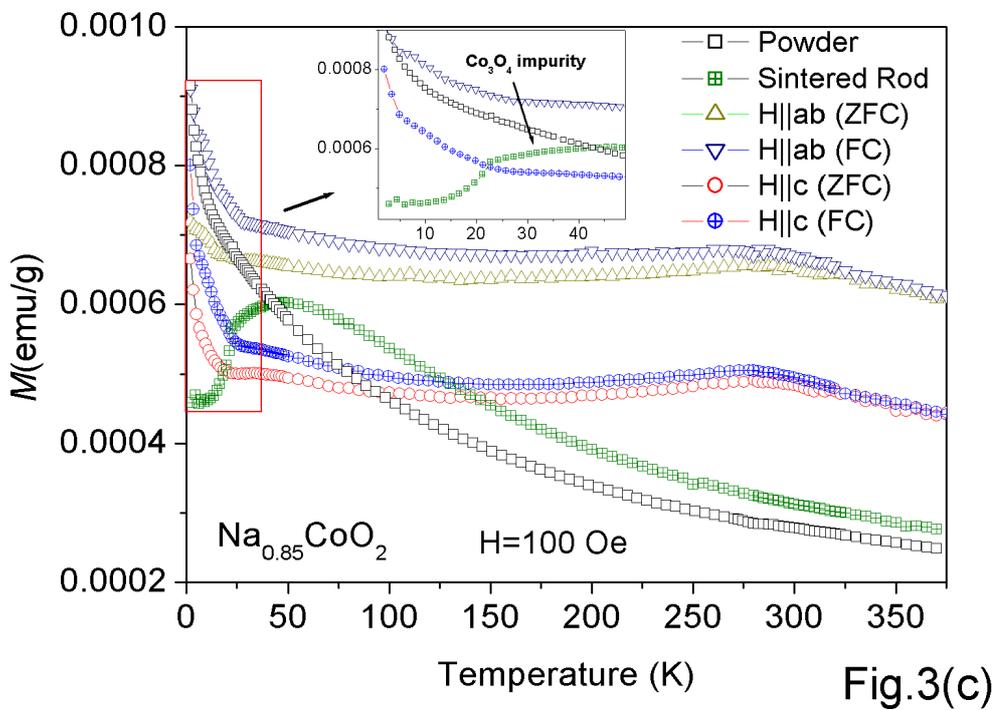

Fig.3(c)

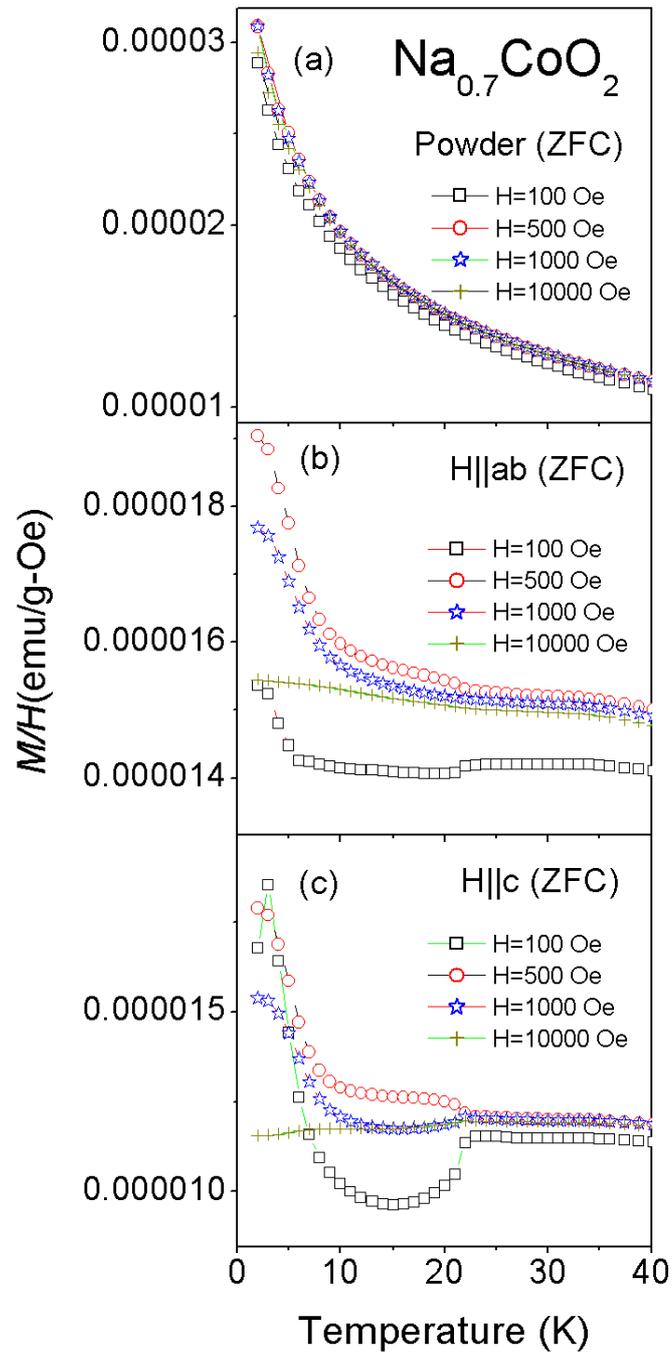

Fig.4

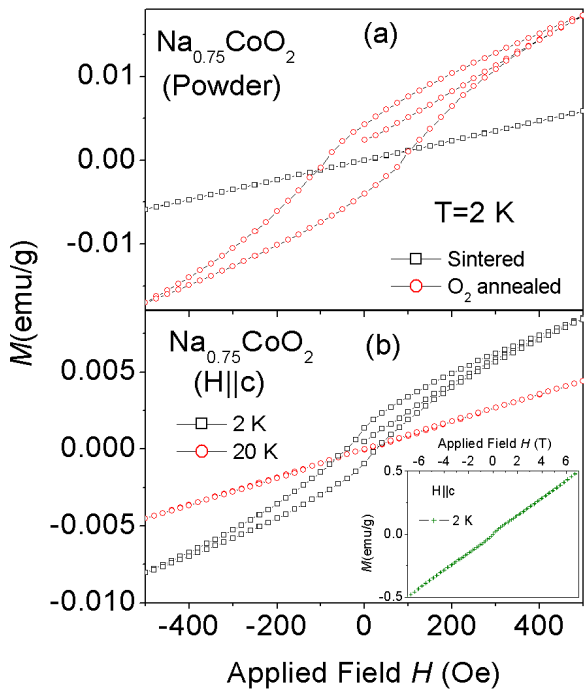

Fig.5

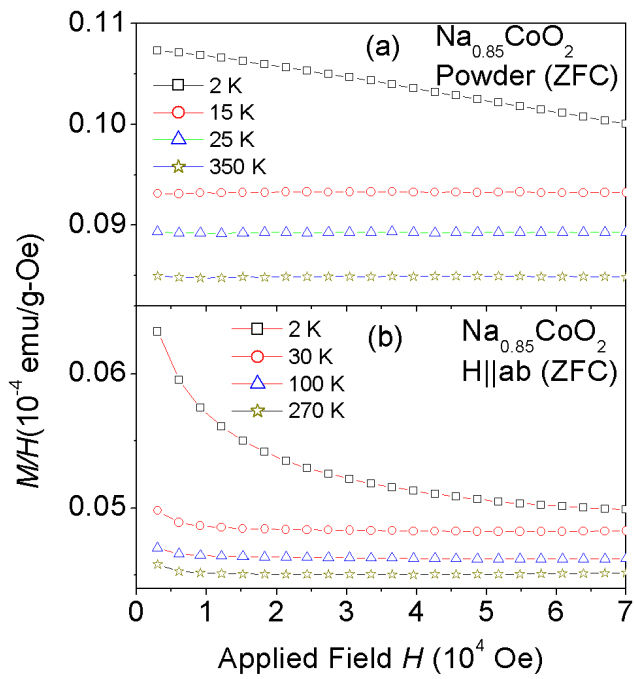

Fig.6